\documentclass[multphys,vecphys]{svmult}

\usepackage{makeidx}     
\usepackage{graphicx}    
\usepackage{multicol}    

\makeindex             

\begin{document}

\title*{Turbulent Structure of the Interstellar Medium}
\author{Mordecai-Mark Mac Low
}
\institute{American Museum of Natural History, 79th Street and Central Park West, New York, NY, 10024-5192, USA
\texttt{mordecai@amnh.org}
}
%
%
\maketitle

\begin{abstract}
How does turbulence\index{turbulence} contribute to the formation and
structure of the dense interstellar medium (ISM)? Molecular clouds are
dense, high-pressure objects.  It is usually argued that gravitational
confinement\index{gravitational confinement} causes the high
pressures, and that the clouds are
(magneto)hydrostatic\index{hydrostatic} objects supported by a balance
between magnetic and turbulent pressures and gravity. However,
magnetic pressures\index{pressure,
magnetic}\index{pressure}\index{pressure, turbulent} appear too weak,
and MHD turbulent support not only requires driving, but also results
in continuing gravitational collapse, as has now been demonstrated in
simulations reaching $512^3$ zones.  
Models of supernova-driven, magnetized turbulence readily form transient,
high-pressure, dense regions 
that may form molecular clouds.  
They are contained not by self-gravity, but by turbulent ram pressures
from the larger flow.  Apparent virialization may actually be a
geometrical effect.  Turbulent clouds are unlikely to be in
hydrostatic equilibrium, instead either collapsing or expanding,
although they may appear well-fit by projected equilibrium
Bonnor-Ebert spheres.  Collapsing clouds probably form stars
efficiently, while expanding ones can still form stars by turbulent
compression, but rather inefficiently.
\end{abstract}

\section{Equilibrium}
\label{sec:equil}

The very high Reynolds numbers of the interstellar medium (ISM) ensure
that it is turbulent and not laminar.  How can we model a turbulent
flow that is intrinsically non-steady and chaotic?  Most models of the
structure of the ISM rely on some notion of equilibrium:
(magneto)hydrostatic equilibrium, virial equilibrium, energy
equipartition, or at least statistical equilibrium (or stationarity).
In this talk I use the structure of the dense ISM as a case study in
the problems with these different assumptions.  In this case, making
such assumptions leads to the conclusion that molecular clouds
are gravitationally bound, which may not generally be the case. 

\section{Standard Argument for Gravitationally Bound Clouds}
\label{sec:standard}
The standard argument assumes that observed clouds are in virial
equilibrium \cite{ms56,m57,cfm99}.  This was justified by apparent
lifetimes estimated at 30 Myr by Blitz \& Shu \cite{bs80}, which is
many times the free-fall time $\tau_{\rm ff}$, and the seeming consistency
of virial and actual mass given size-mass and size-linewidth relations.
Cloud parameters can then be estimated from the assumption of virial
equilibrium, and (magneto)hydrostatic models constructed by detailed
balance of gravity against turbulent and magnetic pressure support.

The full Eulerian virial theorem \cite{mz92} for a volume of space is
a time-dependent equation for the moment of inertia of the mass in the
volume
\begin{equation}
(1/2) \ddot{I} = 2({\cal E}_{th} + {\cal E}_{kin} - {\cal T}_{th} -
{\cal T}_{kin}) + {\cal M} + {\cal W} - (1/2)\dot{\Psi},
\end{equation}
where ${\cal E}_{th}$, ${\cal E}_{kin}$, ${\cal M}$, and ${\cal W}$
are thermal, kinetic, magnetic and gravitational energy, ${\cal
T}_{th}$ and ${\cal T}_{kin}$ are thermal and ram pressure on the
surface of the volume, and $\dot{\Psi}$ is the change in flux of the
moment of inertia through the region.  

This equation is usually simplified by neglecting the time-dependent
and surface terms, and assuming clouds with homogeneous density and
pressure. Below I will address how well justified these assumptions
are, but if they are made, the equation can be recast to give 
external pressure \cite{cfm99}
\begin{equation}
P_{ext} = \frac{1}{4\pi} \left( -\alpha\frac{GM^2}{R^4} +
\beta\frac{\Phi^2}{R^4} + 3\frac{c_s^2 M}{R^3} + \frac{\sigma^2M}{R^3} \right),
\end{equation}
where $M$, $R$, $\sigma$, and $\Phi$ are the mass, radius, velocity
dispersion, and enclosed magnetic flux of a cloud, $G$ is the
gravitational constant, $c_s$ is the sound speed, and $\alpha$ and
$\beta$ are geometrical constants of order unity.  If gravity and
magnetic pressure are ignored, we recover the generalized Boyle's law
$P_{ext}V = (c_s^2 + \sigma^2)M$.  If, on the other hand, external,
thermal, and magnetic pressure are all ignored, the equilibrium is
reduced to a balance between gravity and turbulence:
\begin{equation}
\alpha GM/R = \sigma^2  \Rightarrow M_{vir} = \sigma^2 R / (\alpha G),
\label{mvir}
\end{equation}
which gives the usual definition for the ``virial mass'' $M_{vir}$.
This is certainly appropriate for isolated stellar clusters, where all
the other terms are indeed negligible, but may be rather more
problematic for gas clouds embedded in a turbulent flow.

Clouds will rarely or never have exact equality between gravity and
internal pressure (thermal, magnetic, and turbulent).  McKee
\cite{cfm99} writes the {\em time-averaged} virial theorem in terms of
the mean internal pressure $\bar{P} = P_{ext} - {\cal W} (1 - {\cal
M/|W|})/3V$.  The total energy of the cloud 
\begin{equation}
E = (3V/2)[P_{ext} + {\cal W}(1-{\cal M/|W|})/3V].
\end{equation}
If magnetic fields are unimportant, the cloud is gravitationally bound
if $P_{ext}V < -{\cal W}/3$, that is if the mean weight in virial
equilibrium exceeds the surface pressure.

If we assume the ISM to be in pressure equilibrium, then we can take
the ISM pressure to be the external pressure.  Boulares \& Cox
\cite{bc90} argue for an ISM pressure, neglecting a uniform cosmic ray
component, of $1.8 \times 10^4$~K~cm$^{-3}$.  The gravitational energy
of a cloud with radius $a$ can be expressed in terms of its observable
mean surface density $\Sigma$ or average hydrogen column density
$\bar{N}_H$ as \cite{cfm99}
\begin{equation}
-({\cal W}/3V) = (3\pi a/20) G \Sigma \simeq (1.39 \times 10^5 \mbox{
 K cm}^{-3}) (\bar{N}_H / 10^{22}\mbox{ cm}^{-2}).
\end{equation}
For the local dust-to-gas ratio, $\bar{N}_H = 10^{22}$~cm$^{-2}$
corresponds to a visual extinction of $A_V = 7.5$~mag.  Since
molecular clouds typically have $A_V \gg 2$, $P_{ext} \ll -{\cal
W}/3V$ and so, under the assumptions given, they are gravitationally
bound. 

\section{Validity of Standard Assumptions}

Are the assumptions that go into the standard argument still
well-supported? Let's begin with the assumption that molecular clouds
live for many free-fall times, and therefore have time to virialize.
Blitz \& Shu \cite{bs80} estimated that cloud lifetimes were around
30~Myr in the Milky Way based on three main arguments: the ages of
stars thought to be associated with the clouds; cloud locations
downstream from dust lanes thought to be associated with cloud-forming
spiral-arm shocks; and the comparison between the total molecular mass
and the star-formation rate in the Galaxy.  In the last several years,
much shorter lifetimes have been proposed by Fukui et al.\ \cite{f99}
based on the ages of stellar clusters associated with molecular gas;
and by Ballesteros-Paredes et al.\ \cite{bhv99} based on the lack of a
population of post-T Tauri stars with ages $> 10$~Myr closely
associated with molecular gas.  If molecular cloud lifetimes are only
2--3~$\tau_{\rm ff}$, they may not live long enough to reach virial
equilibrium.

Larson \cite{l81} found that the size of observed molecular clouds
appears to correlate with the linewidth $R \propto \sigma^2$ and with
the mean density of the cloud $R \propto \rho^{-1}$.  These relations
can be interpreted as evidence of virial equilibrium using the
definition of $M_{vir}$ given in equation~(\ref{mvir}), by noting that
if they hold,
\begin{equation}
M_{vir} \propto R \sigma^2 \propto \rho R^3,
\end{equation}
so the virial mass equals the actual mass of the cloud.

\begin{figure}[tbph]
\centering
\includegraphics[width=0.5\textwidth,angle=270]{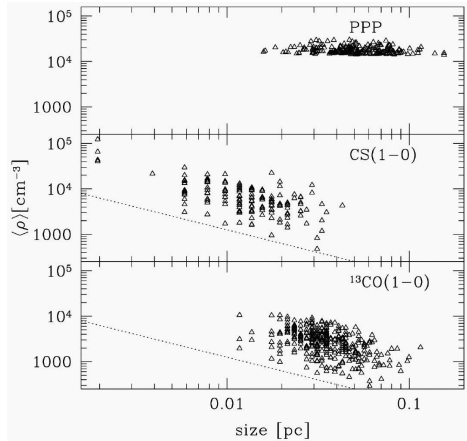}
\caption{Mean density-size relationship for clumps in a turbulent
model measured in physical space (top) and in simulated observational
coordinates for two different species (middle and bottom). The dotted
line has a slope of = -1. In physical space we find no correlation,
verifying the results by VBR \cite{vbr97}, but nevertheless the
simulated observations show such a correlation, as found by Larson
(1981) and many others. The correlation in observational space may
simply be the effect of the limited dynamic range of observations
giving an effective column density cutoff.}
\label{fig:ppp-ppv}       
\end{figure}

Ballesteros-Paredes et al.\ \cite{bvs99} pointed out that this
actually only guarantees that kinetic and gravitational energy are
equal, but not that full virial equilibrium holds, because of the
extra terms neglected in equation~(\ref{mvir}).  Even worse, both
V\'azquez-Semadeni et al.\ \cite{vbr97} and Ballesteros-Paredes \& Mac
Low \cite{bm02} found that simulated observations of turbulent models
appear to reproduce the density-size relationship, but that the actual density
distribution did {\em not}.  Figure~\ref{fig:ppp-ppv} demonstrates this.
Is even the energy equipartition argument merely an observational artifact?

On the other hand, the size-linewidth relationship appears in
turbulent models
\footnote{On a related subject, Brunt \& Mac Low \cite{bm03} examined
the dispersion of velocity centroids in real and simulated
observations.  They found that density inhomogeneity in hypersonic
turbulence cancels projection smoothing, so that observed
two-dimensional velocity centroid scaling actually gives the correct
three-dimensional result, by lucky coincidence.  This is not true in
trans- or sub-sonic turbulence.}  even in the complete {\em absence}
of self-gravity \cite{om02}.  This can be understood as a direct
consequence of the steep velocity power spectrum expected in
turbulence, with index between -5/3 and -2.  Larger objects have
greater velocity differences, and thus larger velocity dispersion.
However, this offers no support for the hypothesis that observed
velocity dispersion are primarily due to self-gravity bringing clouds
to a simple equilbrium between kinetic and gravitational energy.

In summary, the approximation of virial equilibrium for molecular
clouds, especially in its simplest expression of equipartition, may be
misleading.  Their apparent short lifetimes suggest that the transient
surface terms and distortions cannot be averaged away, and may
dominate the dynamics of the cloud.  Ballesteros-Paredes \cite{bvs99}
measured the transient terms in turbulent simulations, and indeed
found them to be substantial.

\section{Clouds in Supernova-Driven Turbulence}

\begin{figure}[tbph]
\centering
\includegraphics[width=1.15\textwidth]{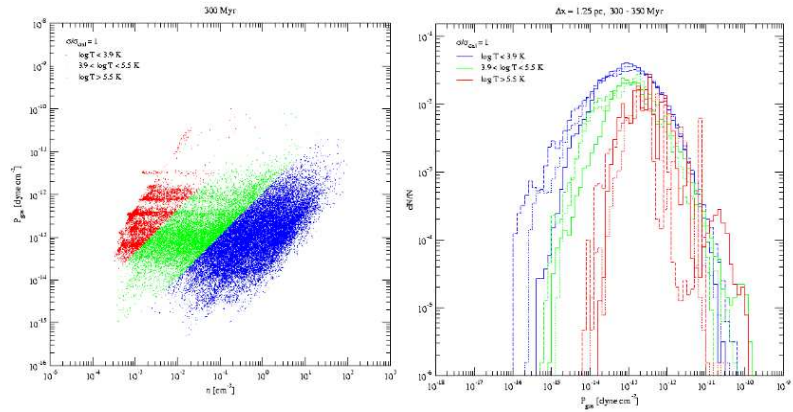}
\vskip -1.8in
\caption{ {\em Left:} Scatter plot of pressure vs.\
density in the midplane of the SN-driven model showing occupation of
high-pressure, high-density region associated with molecular clouds.
{\em Right:} Volume-weighted probability distribution function of
pressure in the same model. Note that the small volumes occupied by
high-density, cold gas have large mass. }
\label{fig:pressures}       
\end{figure}

In order to study the structure of density enhancements in a turbulent
flow more carefully, we turn to a numerical model of supernova-driven
turbulence \cite{a00,am02}, done with an adaptive mesh refinement code
and including: heating initially balancing cooling using an
equilibrium ionization cooling curve; the galactic gravitational
potential (but not self-gravity); and supernovae occurring randomly at
the galactic rate.  In this simulation, large density enhancements
form and dissipate due to the turbulent compression and cooling,
without the participation of self-gravity.  The left side of
Figure~\ref{fig:pressures} shows that pressures in some dense regions
reach values more than an order of magnitude higher than the average,
as observed in molecular clouds.
The pressure probability density function on the right side of the
Figure also shows the broad distribution of pressures.  The
high-density, high-pressure regions are not fully resolved in these
models, but they have sizes, shapes, and masses consistent with giant
molecular clouds.  They {\em must} be transient, probably dispersing
on an internal crossing time \cite{vbk03}, as they cannot be in
equilibrium in the absence of self-gravity.  

If molecular clouds are generally transient, ram-pressure confined
objects, then why do CO luminosity, dust extinction, and other
observational measures of cloud mass correlate so well with apparent
virial mass (see Walter in these proceedings for example)?  As we
noted above, a turbulent flow produces a size-linewidth relation $R
\propto \sigma^2$.  The usual expression for the virial mass,
equation~(\ref{mvir}), can thus be rewritten as $M \propto R^2$.  Any
collection of clouds whose mass is proportional to the square rather
than the cube of their typical size will thus appear to be
``virialized''.  Ram pressure confined objects will usually be
sheetlike, rather than spheres, and will thus naturally have $M
\propto R^2$ rather than $M \propto R^3$.  The filamentary shapes of
observed clouds is also consistent with this idea.  Apparent
virialization of clouds may actually just be a geometrical effect
produced by a compressible turbulent flow.

\section{Hydrostatic Equilibrium}

The assumption of hydrostatic equilibrium between a turbulent pressure
and gravity can also be misleading, because ram pressure is not
isotropic.  Numerical models of self-gravitating, turbulent gas reveal
unexpected behavior.  In the absence of continuing energy input,
turbulence decays quickly \cite{m99}, and collapse proceeds without
substantial delay \cite{kb00}, whether or not fields are present
\cite{bcp01} (so long as they cannot support magnetostatically).  

\begin{figure}[tbph]
\centering
\includegraphics[width=0.4\textwidth,angle=270]{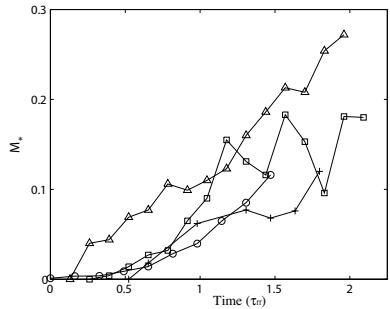}
\caption{Comparison of the mass accretion behavior for runs driven at
$k$ = 1--2 with varying resolution 64$^3$ (triangle), 128$^3$
(square), 256$^3$ (plus), and 512$^3$ (circle) \cite{hmk01,l04}. $M_*$
denotes the sum of masses found in all cores, in units of box mass,
determined by the modified CLUMPFIND \cite{wdb94}.  Collapse rates
vary, but collapse occurs in all cases, with the rate of collapse
generally converging in the higher resolution models.}
\label{fig:accretion}       
\end{figure}

Even driven turbulence strong enough that the turbulent pressure
satisfies the Jeans criterion for stability against gravitational
collapse does not completely prevent local collapse \cite{khm00}.  Adding
weak magnetic fields reduces the amount of collapse, but does not
prevent it entirely \cite{hmk01}, a result that has now been confirmed
at resolutions ranging from $64^3$ zones to $512^3$ zones by Li et
al.\ \cite{l04}, as shown in Figure~\ref{fig:accretion}.
Collapse occurs if the mass in a region exceeds the {\em local} Jeans
mass $m_{J,T} \propto v^3 \rho^{-1/2}$.  Compressible turbulence
produces density enhancements with $\Delta \rho/\rho \propto
v^2/c_s^2$, so even if a region is {\em globally} supported by supersonic
turbulence, {\em local} regions may still become gravitationally
unstable \cite{khm00}.  These results suggest that most observed cores
are dynamically collapsing.

\begin{figure}[tbph]
\centering
\includegraphics[width=0.5\textwidth,angle=270]{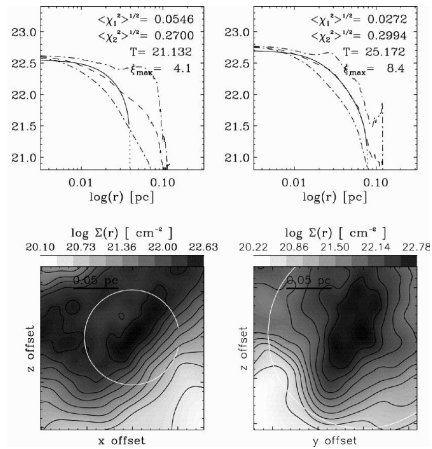}
\caption{Column density maps and radial profiles for xy and xz
projections of a sample clump from an SPH simulation. The white
circles show in each case the size of the radius used in the
Bonnor-Ebert fit. Note the different morphology that the same core
shows in the different projections \cite{bkv03}. 
}
\label{fig:b-e}       
\end{figure}

Observations that appear to suggest cores in hydrostatic equilibrium
also must be interpreted with great care.  More than 50\% of the cores
that appear in a simulated observation of a numerical model of
supersonic, hydrodynamic turbulence {\em without} self-gravity can be
fit well by a projected Bonnor-Ebert sphere \cite{bkv03}, as shown in
Figure~\ref{fig:b-e}.

Gravity clearly does dominate some regions in molecular clouds.
However, gravitational collapse seems to proceed efficiently there,
probably leading to more than half the mass forming stars
\cite{kb01,khm00}.

Large-scale gravitational instability forms regions with masses of
order the thermal Jeans mass of the diffuse ISM $M_J > 10^7 M_{\odot}$
\cite{ko01,kos02}. These instabilities do not directly form single molecular
clouds, but they may form regions of active star formation in which large
molecular clouds form incidentally during gravitational collapse and
fragmentation of the larger region.

\section{Conclusions, or Questions}
Are most or all observed clouds out of equilibrium?  Are they either
ram pressure confined and transient, {\em or} gravitationally
collapsing on a free-fall time?  The observational distinction between
these two states may be whether associated star formation is occurring
on a scale much longer than or close to the free=fall time.  Is
apparently bimodal star formation actually a continuum from low to
high turbulent support?  Finally, do clouds appear to be in virial
equilibrium with $M \propto R\sigma^2$ because they are actually
sheet-like objects in a turbulent flow with $M \propto R^2$ and $R
\propto \sigma^2$?

\vskip 0.1in {\small {\bf Acknowledgments} I thank the organizers for
partial support of my attendance.  My research is partly supported by
NSF grants AST99-85392, and AST03-07793, and NASA grant NAG5-10103.
This work has made use of the NASA ADS Abstract Service.  }
%
%
%

%
%



\printindex
\end{document}